\def\bold#1{\setbox0=\hbox{$#1$}%
     \kern-.025em\copy0\kern-\wd0
     \kern.05em\copy0\kern-\wd0
     \kern-.025em\raise.0433em\box0 }

\def\slash#1{\setbox0=\hbox{$#1$}#1\hskip-\wd0\dimen0=5pt\advance
       \dimen0 by-\ht0\advance\dimen0 by\dp0\lower0.5\dimen0\hbox
         to\wd0{\hss\sl/\/\hss}}

\documentstyle[12pt]{article}

\newlength{\dinwidth}
\newlength{\dinmargin}
\newcommand{\resection}[1]{\setcounter{equation}{0}\section{#1}}

\begin{document}
\vspace*{7cm}
\begin{center}
  \begin{Large}
  \begin{bf}
LIGHT VECTOR  RESONANCES IN THE EFFECTIVE CHIRAL LAGRANGIAN FOR HEAVY
MESONS$^*$\\
  \end{bf}
  \end{Large}
  \vspace{5mm}
  \begin{large}
R. Casalbuoni\\
  \end{large}
Dipartimento di Fisica, Univ. di Firenze\\
I.N.F.N., Sezione di Firenze\\
  \vspace{5mm}
  \begin{large}
A. Deandrea, N. Di Bartolomeo and R. Gatto\\
  \end{large}
D\'epartement de Physique Th\'eorique, Univ. de Gen\`eve\\
  \vspace{5mm}
  \begin{large}
F. Feruglio\\
  \end{large}
Dipartimento di Fisica, Univ.
di Padova\\
I.N.F.N., Sezione di Padova\\
  \vspace{5mm}
  \begin{large}
G. Nardulli\\
  \end{large}
Dipartimento di Fisica, Univ.
di Bari\\
I.N.F.N., Sezione di Bari\\
  \vspace{5mm}
\end{center}
  \vspace{1.5cm}
\begin{center}
UGVA-DPT 1992/07-780\\\
BARI-TH/92-116\\\
July 1992
\end{center}
\vspace{1cm}
\noindent
$^*$ Partially supported by the Swiss National Foundation
\newpage
\thispagestyle{empty}
\begin{quotation}
\vspace*{5cm}
\begin{center}
  \begin{Large}
  \begin{bf}
  ABSTRACT
  \end{bf}
  \end{Large}
\end{center}
  \vspace{5mm}
\noindent
We modify a chiral effective lagrangian recently suggested to describe
interactions of the light pseudoscalars with mesons containing a heavy
quark, so as to incorporate light vector resonances, such as $\rho$,
etc. The modification uses the hidden gauge symmetry approach.
As a preliminary example we present an
application to the semileptonic $D\to K^\star$ decay.
\end{quotation}

\newpage
\setcounter{page}{1}
\def\lq{\left [}
\def\rq{\right ]}
\def\LL{{\cal L}}
\def\VV{{\cal V}}
\def\AA{{\cal A}}

\newcommand{\be}{\begin{equation}}
\newcommand{\ee}{\end{equation}}
\newcommand{\bea}{\begin{eqnarray}}
\newcommand{\eea}{\end{eqnarray}}
\newcommand{\nn}{\nonumber}
\newcommand{\dd}{\displaystyle}

\resection{Introduction}

Much work has been recently devoted to the formulation of an
effective heavy quark theory and to the study of its
predictions \cite{zero}, \cite{Isgur}, \cite{Nussinov}, \cite{Voloshin},
\cite{Bjorken}. Recently Wise \cite{Wise} has proposed an
effective lagrangian to describe at low momentum the interactions of
a meson containing a heavy quark with $\pi$, $K$, $\eta$ (see also
\cite{Yan}, \cite{Lee}). The effective lagrangian possesses the heavy-quark
$SU(2N)$ spin-flavor symmetry (for $N$ heavy quarks) and a
non-linearly realized $SU(3)_L\otimes SU(3)_R$ chiral symmetry in the
light sector, corresponding to spontaneous symmetry breaking
of the chiral group to the diagonal $SU(3)_V$.

The aim of the present paper is to present an extension of such a
description to accommodate inside the scheme also the ``light" vector
resonances as $\rho$, $K^\star$, etc. The effective lagrangian
incorporating the vector
resonances is given below (see eq. (4.3)). We will then apply the
formalism, in a preliminary way, to the decay $D\to K^\star\ell\nu$.

To introduce vector resonances we will follow the standard hidden gauge
symmetry technique already used in previous cases \cite{Bala}, \cite{Bando},
\cite{Bess}.

A further extension of the method allowing for the introduction of
axial-vector
resonances has been described in \cite{Bando} and in  \cite{assiali}.
For the applications described in this paper we will consider only the
case of the octet of vector resonances, and we will not discuss the
also known
enlargement to $U(3)_L\otimes U(3)_R$, that would be necessary to study
the nonet, besides the extension to the axial-vector case.

\resection{Non linearly realized chiral symmetry}

Let us recall the basic elements leading to the effective lagrangian of
Wise \cite{Wise}.

The description of a light pseudoscalar meson can be encoded in the
$3\times 3$ unitary matrix
\be
\Sigma=\exp{\frac{2iM}{f}}
\ee
where $f\approx 132~MeV$ is the pseudoscalar pion decay constant
and $M$ is the matrix
\be
{M}=
\left (\begin{array}{ccc}
\sqrt{\frac{1}{2}}\pi^0+\sqrt{\frac{1}{6}}\eta & \pi^+ & K^+\nn\\
\pi^- & -\sqrt{\frac{1}{2}}\pi^0+\sqrt{\frac{1}{6}}\eta & K^0\\
K^- & {\bar K}^0 &-\sqrt{\frac{2}{3}}\eta
\end{array}\right )
\ee
At the lowest order in the derivatives, and in the massless limit, the
chiral lagrangian is
\be
\LL_{\rm chiral}=\frac{f^2}{8}Tr\left[\partial^\mu\Sigma\partial_\mu
\Sigma^\dagger\right]
\ee
showing the invariance under the transformations of
$SU(3)_L\otimes SU(3)_R$:
\be
\Sigma\to g_L\Sigma {g_R}^\dagger
\ee
To describe interactions with other fields it turns out convenient
to go to the CCWZ realization \cite{CCWZ}:
\be
\Sigma=\xi^2
\ee
with $\xi$ transforming in the following way
\be
\xi\to g_L\xi U^\dagger=U\xi g_R^\dagger
\ee
where $U\in SU(3)$ is a non-linear function of $g_L$, $g_R$ and $\xi$.

The effective description
of the ground state of the system $Q{\bar q}^a$ ($Q$ is $c$ or $b$
and $q^1=u$, $q^2=d$,
$q^3=s$) uses the $4\times 4$ matrix $H_a$ \cite{Georgi}, \cite{Wise}
($a=1,2,3$)
\be
H_a=\frac{(1+\slash v)}{2}[P_{a\mu}^\star\gamma^\mu-P_a\gamma_5]
\ee
which transforms under $SU(3)_L\otimes SU(3)_R$ as
\be
H_a\to H_b U^\dagger_{ba}(x)
\ee
We recall that in (2.7) $P_{a\mu}^\star$ and $P_a$ annihilate
respectively a spin-$1$ and spin zero meson $Q{\bar q}_a$ of
velocity $v_\mu$ ($v^\mu P_{a\mu}^\star=0$). The heavy quark spin
symmetry $SU(2)_{v}$ acts as
\be
H_a\to SH_a
\ee
 for $S\in SU(2)_{v}$, satisfying
$[\slash v,S]=0$, and under Lorentz transformations
\be
H_a\to D(\Lambda)H_aD(\Lambda)^{-1}
\ee
Let us also recall that one defines
\be
{\bar H}_a=\gamma_0 H_a^\dagger \gamma_0
\ee

\resection{Description through a linear gauge symmetry}

Eq. (2.6) implies in particular, that the transformation $U$
depends on the space-time.
It is then possible to get a simpler, linear realization, by using the
{\underbar{hidden}} {\underbar{gauge}} {\underbar{symmetry} approach.
This consists in using two new $SU(3)$-matrix valued fields $L$ and $R$
to build up $\Sigma$
\be
\Sigma=LR^\dagger
\ee
The chiral lagrangian in eq. (2.3) is then invariant under the
group $SU(3)_L\otimes SU(3)_R\otimes SU(3)_H$
\be
L\to g_L L h^\dagger(x),~~~~~~~R\to g_R R h^\dagger(x)
\ee
where $h\in SU(3)_H$ is a local gauge transformation. The local symmetry
associated to the group $SU(3)_H$ is called the
hidden gauge symmetry because the
field $\Sigma$ belongs to the singlet representation. It should be
noticed that this description is equivalent to the previous one by
the gauge choice $L=R^\dagger$, which is always possible to make (at
least locally). {}From the fields $L$ and $R$ we can construct two
currents
\be
\VV_\mu=\frac{1}{2}\left(L^\dagger\partial_\mu L+R^\dagger
\partial_\mu R\right)
\ee
\be
\AA_\mu=\frac{1}{2}\left(L^\dagger\partial_\mu L-R^\dagger
\partial_\mu R\right)
\ee
which are singlets under $SU(3)_L\otimes SU(3)_R$ and transform as
\be
\VV_\mu\to h\VV_\mu h^\dagger+h\partial_\mu h^\dagger
\ee
\be
\AA_\mu\to h\AA_\mu h^\dagger
\ee
under the local group $SU(3)_H$.

In this notation, the transformation for $H_a$ reads
\be
H_a\to H_b h_{ba}^\dagger(x)
\ee
and the covariant derivative is defined as
\be
D_\mu{\bar H}=(\partial_\mu+\VV_\mu){\bar H}
\ee

\resection{Inclusion of vector mesons}

The octet of vector resonances ($\rho$, etc.) are introduced as the gauge
particles associated to the group $SU(3)_H$. We put
\be
\rho_\mu=i\frac{g_V}{\sqrt{2}}\hat\rho_\mu
\ee
where $\hat\rho$ is a hermitian $3\times 3$ matrices analogous to the one
defined in equation (2.2). This field transforms under the full symmetry
group as $\VV_\mu$
\be
\rho_\mu\to h\rho_\mu h^\dagger+h\partial_\mu h^\dagger
\ee
The vector particles
acquire a common mass through the breaking of
$SU(3)_L\otimes SU(3)_R\otimes SU(3)_H$ to $SU(3)_V$.
In fact, 8 out of the
16 Goldstone bosons coming from the breaking are the light pseudoscalar
mesons, whereas the other 8 are eaten up by the $\rho$ field.

{}From all the preceding transformation laws one is lead to the
following simple lagrangian which incorporates the Wise lagrangian
and the vector-octet interactions with the heavy mesons and the
pseudoscalars
\bea
\LL&=&\LL^{\rm light}+iTr[H_a v_\mu\partial^\mu {\bar H}_a]
+i Tr[H_b v^\mu(\VV_\mu)_{ba}
{\bar H}_a]\nn\\
& &+ig Tr[H_b\gamma_\mu\gamma_5(\AA_\mu)_{ba}{\bar H}_a]+
i\beta Tr[H_bv^\mu\left(\VV_\mu-\rho_\mu\right)_{ba}{\bar H}_a]\nn\\
& &+\frac{\beta^2}{2f^2 a}Tr[{\bar H}_b H_a{\bar H}_a H_b]\nn\\
& &+\lambda_0 Tr[m_q\Sigma+m_q\Sigma^\dagger]\nn\\
& &+\lambda_1 Tr[{\bar H}_a(R^\dagger m_q L+L^\dagger m_q R)_{ab}
{\bar H}_b]\nn\\
& &+\lambda_1^\prime Tr[H_a{\bar H}_a
(m_q\Sigma+\Sigma^\dagger m_q)_{bb}+
\cdots
\eea
where the ellipsis denotes terms with additional derivatives as well
as higher order mass corrections ($1/m_Q$, etc.),
and $\LL^{\rm light}$ is
\be
\LL^{\rm light}=-\frac{f^2}{2}\left\{tr[\AA_\mu \AA^\mu]+a tr[(\VV_\mu-
\rho_\mu)^2]\right\}+\frac{1}{2g_V^2}tr[F_{\mu\nu}(\rho)F^{\mu\nu}(\rho)]
\ee
In eqs. (4.3) and (4.4) $g$ (the same one occurring in the Wise
lagrangian),
$\beta$, $a$ are constants ($f$ and $g_V$ had been defined above),
\be
F_{\mu\nu}(\rho)=\partial_\mu\rho_\nu-\partial_\nu\rho_\mu+
[\rho_\mu,\rho_\nu]
\ee
the trace operations are $Tr$ on spinor indices and $tr$ on group
indices. In (4.3) a sum on the velocities $v_\mu$ is understood.

The quartic
term in the heavy meson fields in (4.3) follows from the requirement that
the lagrangian goes back to the Wise lagrangian in the formal limit
$g_V\to\infty$ ($m_\rho\to\infty$) in which the $\rho$ field decouples.

The lagrangian $\LL^{\rm light}$, eq. (4.4), describes light pseudoscalars
and vector resonances. It reproduces in this sector all the good results
of vector dominance \cite{Bando}. Of the three terms in eq. (4.4), the
first one reproduces the chiral lagrangian of eq. (2.3),
as it can be seen by using the identity
\be
Tr[\AA_\mu \AA^\mu]=-\frac{1}{4}Tr[\partial_\mu\Sigma\partial^\mu
\Sigma^\dagger]
\ee
By coupling to the electromagnetic field one gets that the first KSRF
relation
\cite{KSRF} is automatically satisfied, i.e. $g_\rho=g_{\rho\pi\pi} f^2$,
where $g_\rho$ is the $\rho-\gamma$ mixing. Furthermore by the choice $a
=2$ it is also possible to satisfy the second KSRF relation, $m_\rho^2=
g_{\rho\pi\pi}^2f^2$. Using (4.3) we get
\be
m_\rho^2=\frac{1}{2} a g_V^2f^2
\ee
and for $a=2$
\be
g_V=\frac{m_\rho}{f}\approx 5.8
\ee
Finally, the light-quark mass-terms in eq. (4.3) are the same as
defined by Wise.

Eq. (4.3) is the lagrangian we propose as an attempt to improve on the
chiral Wise lagrangian by taking into account also light vector mesons
effects. The extension to include axial-vector resonances can easily be
made following references \cite{Bando} and \cite{assiali}.
Also, one can easily enlarge to $U(3)_L\otimes U(3)_R$ \cite{Bando},
necessary to study the full resonance nonets. As an example, we will
present in the next section a preliminary, and as we will see
probably incomplete, discussion of the semileptonic decay
$D\to K^\star\ell\nu$, which does not require at this stage the
extensions we have mentioned.

\resection{Application to the semileptonic decay ${D \to K^\star}$}

We will apply the previous formalism to the calculation of semileptonic
decays of $B$ and $D$ mesons to ``light" vector states as $\rho$ and
$K^\star$. We will consider the explicit calculation of $D^+\to
{\bar K}^{0\star}\ell\nu$.

First of all we need to construct an effective
current between heavy mesons and light vectors transforming as
$({\bar 3}_L,1_R)$ under $SU(3)_L\otimes SU(3)_R$ (remember that the
quark current is $J_a^\mu={\bar q}_a\gamma^\mu(1-\gamma_5)Q)$. The lowest
dimension relevant operator is
\be
L_a^\mu=i\frac{\alpha}{2}Tr[\gamma^\mu(1-\gamma_5)H_b\xi^\dagger_{ba}]+
\alpha_1Tr[\gamma_5H_b(\rho^\mu-\VV^\mu)_{bc}\xi_{ca}^\dagger]+\cdots
\ee
where the ellipsis denotes terms vanishing in the limit $m_q\to 0$,
$m_Q\to\infty$ or terms with derivatives. The
constant $\alpha$ can be obtained by considering the matrix element of
$L_a^\mu$ between the meson state and the vacuum, with the result
\be
\langle 0|{\bar q}\gamma^\mu\gamma_5 Q| P\rangle=if_Pm_Pv^\mu
\ee
In the limit $m_Q\to\infty$
\be
\alpha=f_P\sqrt{m_P}
\ee
and $\alpha$ has a calculable logarithmic dependence on the heavy quark
masses.

For the following we will need also the parametrization of the matrix
element of the quark current in terms of form factors
\bea
\langle {\bar K}^{0\star}(p^\prime,\epsilon)|J_\mu|D^+(p)\rangle &=&
\epsilon_{\mu\nu\rho\sigma}\epsilon^{*\nu}p^\rho {p^\prime}^\sigma
\frac{2V(q^2)}{(M_D+M_{K^\star})}+i\epsilon_\mu^*(M_D+M_{K^\star})
A_1(q^2)\nn\\
& & -i(\epsilon^*\cdot q)\left[(p+p^\prime)_\mu\frac{A_2(q^2)}
{(M_D+M_{K^\star})}+2\frac{M_{K^\star}}{q^2}q_\mu A(q^2)\right]
\eea
where $q=p-p^\prime$ and $A(0)=0$. Using the effective lagrangian given
in eq. (4.3) we find two contributions to the calculation of the matrix
element of the current. The first one (see Fig. 1) is just the matrix
element of $L^\mu$ between meson states, and it gives
\be
-2i\frac{g_V}{\sqrt{2}}\alpha_1\sqrt{M_D}\epsilon_\mu^*
\ee
Then we have the contribution from the exchange of $D_s^+$ (see Fig. 2).
The trilinear coupling between $D_s$, $D$ and $K^\star$ can be easily
derived from the last term in the lagrangian (4.3). We find
\be
-i\beta\frac{g_V}{\sqrt{2}}
\frac{f_D(\epsilon^*\cdot q)}{M_D(v\cdot p^\prime+\Delta^\prime)}q_\mu
\ee
where $\Delta^\prime=M_{D_s}-M_D$, $v_\mu$ is the four-velocity of $D_s$,
and $f_D=(\alpha/\sqrt{M_D})$ is the decay coupling constant of the meson
$D$ (see \cite{Wise}). The mass splittings among the heavy mesons are
treated by adding convenient terms to the lagrangian (4.3) (a color
magnetic moment operator). This is discussed in
ref. \cite{Wise} together with the modifications induced on the heavy
quark propagators.

We see that the lagrangian (4.3) gives contributions only to the form
factors $A_1(q^2)$ and $A(q^2)$. However there are other possible pole
terms
that could contribute. For instance the ones coming from the p-wave
meson excitations. The necessary formalism to deal with these states has
been discussed by Falk \cite{Falk}, but, at least in this paper, we will
not consider these contributions, our main motivation being to
illustrate the formalism, rather than to perform a complete
calculation. However, there is another possible contribution that we can
take into account, that is the pole contribution from a $D_s^\star$ meson
as illustrated in Fig. 3. The lagrangian (4.3)
does not contribute to the trilinear coupling among $D$, $D_s^\star$ and
$K^\star$, so we need to consider a higher dimensional invariant
operator to add to it
\be
i\lambda Tr[H_a\sigma_{\mu\nu}F^{\mu\nu}_{ab}(\rho){\bar H}_b]
\ee
The contribution from the diagram of Fig. 3 is then
\be
-\frac{g_V}{\sqrt{2}}
\frac{2\lambda f_D}{v\cdot p^\prime+\Delta}\epsilon_{\mu\nu\rho\sigma}
\epsilon^{*\nu} p^\rho p^{\prime\sigma}
\ee
with $\Delta=M_{D_s^\star}-M_D$.

Summarizing, we get the following
expressions for the form factors defined in eq. (5.4) at $q^2_{max}$
\bea
V(q^2_{max}) &=& -\frac{g_V}{\sqrt{2}}
\frac{\lambda f_D(M_D+M_{K^\star)}}{M_{K^\star}+\Delta}\nn\\
A_1(q^2_{max}) &=&-\frac{g_V}{\sqrt{2}}
\frac{2\alpha_1\sqrt{M_D}}{(M_D+M_{K^\star})}\nn\\
A_2(q^2_{max}) &=& 0\nn\\
A(q^2_{max}) &=& \frac{g_V}{\sqrt{2}}\frac{\beta f_D}
{M_D(M_{K^\star}+
\Delta^\prime)}\frac {(M_D - M_{K^\star})^2}{2M_{K^\star}}
\eea
As we have discussed previously, the reason why
we find $A_1$ independent of $q^2$ and $A_2=0$, is that we are
neglecting the p-wave contributions. However these states do not
contribute to the form factor $V(q^2)$.

We expect our calculation to be reliable
only around the region of maximum recoil $q^2=(M_D-M_{K^\star})^2$, so
in order to compare with experimental data and with other models we
need to extrapolate our form factor down to $q^2=0$.
For this, we assume the pole
dominance of $V$ from the $D_s^\star$ exchange
\be
V(q^2)=V(0)\frac{M_{D_s^\star}^2}{M_{D_s^\star}^2-q^2}
\ee
The expression we get for $V(0)$ is
\be
V(0)=-\frac{g_V}{\sqrt{2}}
\frac{\lambda f_D (M_D+M_{K^\star})(2M_D-M_{K^\star}+\Delta)}
{M_{D_s^\star}^2}
\ee
An evaluation for $\lambda$ can be obtained using the recent data
{}from the E653 Collaboration \cite{E653}
\be
V(0)=0.99\pm 0.27
\ee
We get (using $f_D\approx 200~MeV$ as obtained from QCD sum
rules \cite{QCD})
\be
\lambda=-(0.63\pm 0.17)~GeV^{-1}
\ee
This number turns out to be of the right order of magnitude, as we
expect that the momentum expansion is regulated by the chiral symmetry
breaking scale, which is believed to be around $1~GeV$.

Unfortunately,
{}from this process, it is impossible to draw any conclusion about the
$\beta$ parameter, because the form factor $A(q^2)$ is practically
not accessible to the experiments. We may have a further check on
our approach by studying the form factor $V$ for the companion
semileptonic processes $D\to \rho$ and $B\to\rho$. We can easily evaluate
these processes following the same steps as for $D\to K^\star$,
with obvious changes (for example $D_s^\star\to D^\star$ and
$D_s^{\star}\to B^{\star}$ respectively, $M_{K^\star}\to M_{\rho}$, and
$\Delta=46~MeV$ for $B\to\rho$).

However,
due to the lack of experimental data about these reactions we can only
compare our results with the ones obtained in other calculations.
A comparison is shown in the following table,
for $V(q^2_{max})\equiv V^m$ in the various processes
assuming as normalization $D\to K^\star$, and using the scaling relation
$f_B/f_D=\sqrt{(m_D/m_B)}$.

\begin{center}
\begin{tabular}{|c|c|c|c|c|}
\hline
&Our result&WSB \cite{Steck}&AW \cite{Quark}&LMMS
\cite{Lattice}\\
\hline
$
V^m_{D\to\rho}/V^m_{D\to K^\star}$
&1.25 &1.05 &1.20 & 0.98 \\
\hline
$
V^m_{B\to\rho}/V^m_{D\to K^\star}$
&1.90 &0.73 &1.64 &-\\
\hline
\end{tabular}
\end{center}

Note however that the above scaling law from the meson decay constants is
most probably afflicted by order ${\cal O}(1/m_Q)$ corrections, presumably
still big at the charm mass. This appears from studies of QCD sum rules
\cite{QCD}, and lattice QCD calculations \cite{Allton}.

In fact, besides the non inclusion of the p-waves excitations in the
calculations, that we have already mentioned, the lack of a discussion
of the non-leading contributions in the derivative expansion and of
neglected ${\cal O}(1/m_Q)$ corrections is one of the weakest points
of the present calculation, which we have presented mostly as a
concrete example of application of the effective lagrangian (4.3).
The fact that the $K^\star$ mass is not negligibly small as compared
to the $D$ mass leads additional work as inevitable before coming to
final conclusions.

\resection{Conclusions}

The lagrangian in eq. (4.3) is proposed as a convenient way to incorporate
light vector resonances, such as $\rho$, $K^\star$, etc., into the effective
lagrangian for mesons containing a heavy quark. We have applied such a
lagrangian to the $D\to K^\star$ semileptonic decay, to see how the formalism
may work in a particular case but leaving apart the discussion of
non-dominant effects.

\vspace{1cm}
\noindent
{\bf ACKNOWLEDGMENT}: We would like to thank S. De Curtis and D. Dominici
for many fruitful and enlightening discussions during
the early stage of this work.

\newpage
\begin{center}
  \begin{Large}
  \begin{bf}
  Figure Captions
  \end{bf}
  \end{Large}
\end{center}
  \vspace{5mm}
\begin{description}
\item [Fig. 1] Feynman diagram giving rise to the contribution (5.5).
The shaded square denotes the current of eq. (5.1).
\item [Fig. 2] Feynman diagram giving rise to the contribution (5.6).
The shaded square denotes the current of eq. (5.1).
\item [Fig. 3] Feynman diagram giving rise to the contribution (5.8).
The shaded square denotes the current of eq. (5.1).
\end{description}

\end{document}